\documentclass[runningheads]{llncs}
\usepackage[T1]{fontenc}
\usepackage{graphicx}
\usepackage{subcaption}

\begin{document}
\title{SPARC: Shared Perspective with Avatar Distortion for Remote Collaboration in VR}
\titlerunning{SPARC: Shared Perspective with Avatar Distortion}
%
\author{João Simões\inst{1} \and
Anderson Maciel\inst{1,2}\orcidID{0000-0002-0780-6555} \and
Catarina Moreira\inst{2,3}\orcidID{0000-0002-8826-5163}\and
Maurício Sousa\inst{4}\orcidID{0000-0003-1438-2882}
\and
Joaquim Jorge\inst{1,2}\orcidID{0000-0001-5441-4637}}
\authorrunning{J. Simões et al.}
%
\institute{Instituto Superior Técnico, University of Lisbon, Portugal \and
INESC-ID, Lisbon, Portugal
\email{\{anderson.maciel, jorgej\}@tecnico.ulisboa.pt}
\and
University of Technology Sydney, Australia\\
\email{catarina.pintomoreira@uts.edu.au}
\and
University of Toronto, Canada\\
\email{mauricio.sousa@utoronto.ca}
}%
\maketitle              
\begin{abstract}

We introduce SPARC, a new approach to shared perspective in multi-user collaboration. Each user has a first-person view of the workspace while seeing the others' avatars at their positions around the workspace. Whenever a user manipulates an object, others will see his/her arms stretching to reach that object in their first-person perspective. SPARC combines shared orientation and support to non-verbal communication, minimizing occlusions. We conducted a user study (n=18) to understand how the novel approach impacts task performance and workspace awareness. We found evidence that SPARC is more efficient and less mentally demanding than life-like settings.
\keywords{Collaborative UI \and Virtual Reality \and Shared Perspective \and Avatar Manipulation}
\end{abstract}
\section{Introduction}
\label{chap:intro}
Remote work has increased quickly with the recent pandemic and is carried out primarily through the many available teleconferencing solutions. However, videoconferencing systems limit users' interactions with counterproductive communication layers. Even the most widely used teleconferencing solutions cannot avoid cumbersome steps to what would be easily performed in a face-to-face meeting~\cite{10.1145/1128923.1128947}, for example, various people pointing to a diagram in a presentation. These communication barriers are caused by disregarding how users communicate in a face-to-face setting, including gestures, body posture, awareness of where others are looking, and side conversations.

Past works have shown that using Virtual Reality (VR) is a better approach to cooperation between multiple users, as it provides new but natural ways of visualizing and interacting with objects, the environment and people~\cite{10.1145/293701.293715,6790594,10.1145/1128923.1128947}. This is especially useful when there is a need to manipulate and discuss objects such as 3D models, which is a very common setting in many fields, such as engineering, architecture, geophysical exploration, and even medicine. Effective remote collaboration requires users to understand each other and know where others are in the environment and what they are doing unambiguously.

Current VR technology can simulate face-to-face settings faithfully. However, when collaborating over shared content, people have different points of view on the environment and the task at hand. Despite being natural, this disparity can induce some ambiguities with opposing orientations, i.e., left and right~\cite{10.1145/3242587.3242594}, which is especially true when many collaborators are working around a table. While this also occurs in physical environments, VR can simulate environments one "step outside of the normal bounds of reality"~\cite{10.3389/frobt.2016.00074}, enhancing it. 

Thus, a virtual environment (VE) can be designed to purposefully manipulate individual perspectives into a shared perspective between users~\cite{MAGIC,10.1145/3359997.3365744}. It is as if all users were seated on the same chair, seeing the work area from the same point of view (POV), with the difference that additional pairs of hands are visible, and each user controls their pair. Shared perspective has induced a more effective collaboration between two users~\cite{10.1145/3173574.3173855,7563559}. However, it is limited as it removes the non-verbal communication offered by a face-to-face setting. Not addressing the unsolved issues can hurt collaborative efforts, hinder the awareness of others in collaborative tasks, and interfere with the perceived workload and the social presence felt by the collaborators. Moreover, when there is a need for multiple experts in different fields, more than two users must be supported. Using the current approaches with multiple users would cause the visual space to become cluttered, a reason why shared perspectives with multiple users have not been explored. 

\begin{figure}[t]
  \includegraphics[width=0.98\linewidth]{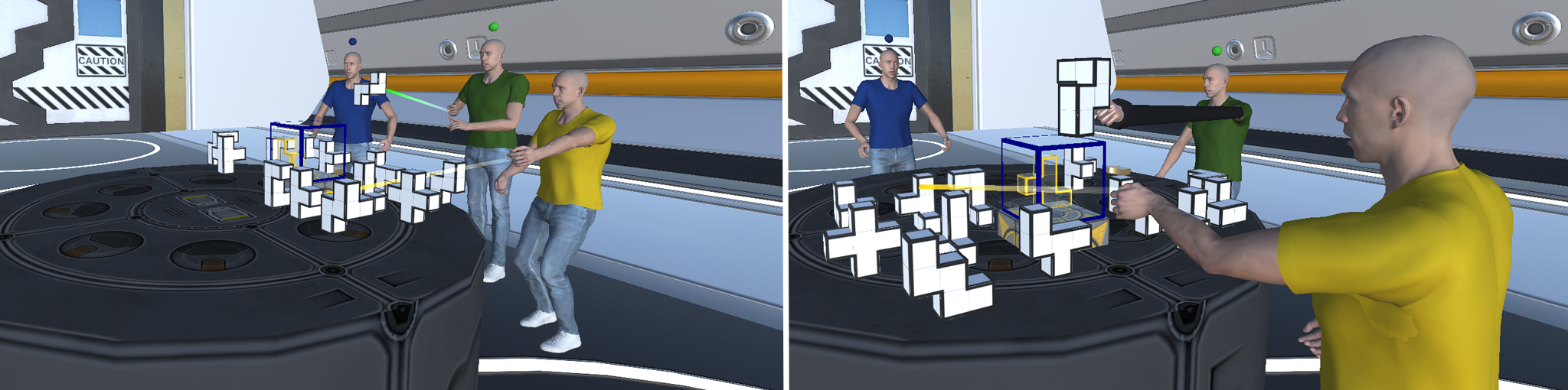}
  \caption{Over-the-shoulder collaboration (top) is replaced by shared perspective with stretched avatars (bottom). While the perspective is shared, the avatar visualization is face-to-face, allowing for non-verbal cues.}
  \label{fig:teaser}
\end{figure}

This paper introduces a solution to cluttering and lack of face-to-face non-verbal communication. With our approach, multiple users work around a table in a VE. They can see each other in their respective seats at the same time that they all see the working objects from the same perspective (Fig.~\ref{fig:teaser}). It works as if there were multiple synchronized copies of the scenario, one before each user. Nevertheless, this is not enough because when you point to an object in front of you, your co-workers will not see where your hand is pointing in their copy of the scene. Therefore, we present a solution based on manipulating the user representations (avatars) that provides a faithful localization of references in the workspace. This approach stretches the users' arms representation in the others' environments while maintaining the local user in their virtual body. In the remainder of the paper, we detail the technique and present an experimental study we conducted on the impact of the method on metrics such as task performance, the feeling of co-presence, and workspace awareness (WA) compared to a conventional VR setting without a shared perspective. 

\section{Related Work}
\label{relatedWork}


Awareness of groupware is a common topic in the CSCW community~\cite{greenberg1996awareness,10.1145/257089.257284}. The lack of awareness is the reason for the usual feeling of inefficiency and awkwardness felt in many groupware systems compared to face-to-face work~\cite{10.1145/257089.257284}. Gutwin and Greenberg~\cite{gutwin2002descriptive} suggest that in a groupware system, awareness information should be presented in a natural way, as people have grown accustomed to co-located environments, by making use of the following workspace awareness (WA) gathering mechanisms:  \textit{consequential communication}, \textit{feedthrough} and  \textit{intentional communication}. Although these mechanisms happen naturally in a co-located environment, maintaining WA is challenging in a remote collaboration system with current technological limitations.

Ishii et al.\cite{10.1145/142750.142977} discussed three metaphors to integrate person and task space: the whiteboard metaphor, the over-the-table metaphor, and the glass window metaphor. With these metaphors, there is a clear distinction between person and task space, so collaborators must shift focus from the task space to read others’ body language and nonverbal cues, which can be straining and hinder communication efforts. Other approaches propose alternatives for face-to-face settings~\cite{10.1145/2642918.2647393} or separate person and task spaces but closely couple task and reference spaces~\cite{10.1145/2207676.2208333}. 
Sodhi et al.~\cite{10.1145/2470654.2470679} developed BeThere, an AR solution that renders the remote user's hand in the local user's workspace. This approach supports intentional communication and feedthrough, although non-verbal communication cues are absent due to only the person's hands being shown. 

In many of the previous cases, orientation was inverted among users. Inversion of orientation is a downside, which can be solved by mirror-reversing one of the participants~\cite{gutwin2002descriptive}. Piumsomboon et al.~\cite{10.1145/3173574.3173620} found that being able always to keep communication cues in sight of the collaborators improved collaborative task completion times and reduced mental effort. Fussel et al.~\cite{fussell2004gestures} argued that performance on collaborative tasks and communication was enhanced when participants had a more complete perception of the gestures the other collaborators were executing. 
Sousa et al.~\cite{10.1145/3359997.3365744} argued that sharing the same perspective decreased mental effort, whereas Medeiros et al. argue that a first-person perspective is paramount~\cite{medeiros2018keep}. Hoppe et al.~\cite{10.1145/3432950} obtained similar results where intentional communication and shared a common perspective enhanced task efficiency. However, spatial inconsistency between where the user is looking and where the user’s avatar is might confuse and break the feeling of presence. Fidalgo et al.~\cite{MAGIC} employed a shared perspective over the task space with manipulations to users’ avatars to preserve deictic references without them being noticeable by the second user. It was argued that there was an improvement in WA and in the feeling of co-presence.

In summary, WA requires the three types of communication to be successfully maintained~\cite{10.1145/2642918.2647393,10.1145/3359998.3369404,MAGIC,10.1145/142750.142977,LI201723,4480795}. However, few references support manipulation of 3D virtual objects or multiple users. Approaches that combine all of the different types of communication are scarce. To our knowledge, ShiSha~\cite{10.1145/3432950} is the closest approach to doing it, since avatar and perspective manipulations are also employed to provide a shared perspective in a multi-user setting. However, the authors do this through the disembodiment of the avatar, which hinders consequential communication, since users may not be aware of others' body language cues. 

\section{SPARC Approach}
\label{approach}

\begin{figure}[t]
    \centering
    \includegraphics[width=1\linewidth]{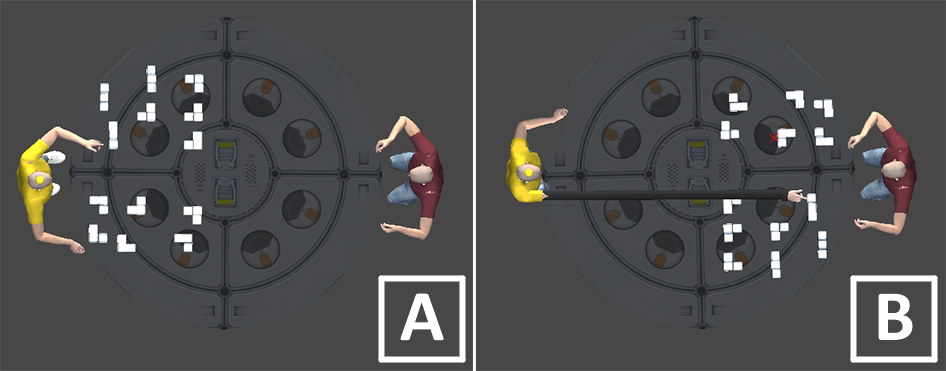}
    \caption{Avatar manipulation: [A] - User on the left points at a piece in his local workspace; [B] - User on the right sees this reference mapped to local own workspace }
    \label{fig:avatar_manipulation_final}
\end{figure}

Here we introduce SPARC, a multi-user VR collaborative technique that enables users to work together from the same perspective while maintaining face-to-face communication, distorting the Avatar Representation as needed to reconcile face-to-face communication with a shared perspective (Fig.~\ref{fig:avatar_manipulation_final}). We developed a Proof of Concept (POC) using Unity, incorporating Photon Engine's PUN  multiplayer networking engine. It provides support for various VR platforms, including the Meta Quest 2 used in our prototype.

\subsection{Improving Workspace Awareness}
We an over-the-table setting, where users sit around a table with a collaborative task between them. 
The task space is the table placed between the users, which will also correspond to the reference space where they can use gesturing and pointing to reference the work. Additionally, users can see each other from across the table, which simplifies capturing the nonverbal cues that happen in person space. This ensures consequential and intentional communication since users can see each other face-to-face, thus understanding where the other is looking, their position, and their actions.

SPARC also enables users to collaborate over one or more objects in the workspace, allowing for differences in their actions to be noticeable to all participants to ensure feedthrough. 

While in the classic over-the-table metaphor, the perspective will be inverted for one of the users in a pairwise collaboration~\cite{10.1145/142750.142977}, we provide the same shared perspective for all users. This also removes the problem of the occlusion of gestures by the objects in the workspace. To preserve the meaning of gestures, transformations are applied to map the user's intended referencing position to the others' work-space, i.e., if the user points to a piece on the right of their workspace, the other users will see the avatar's arm stretched and pointing to the right of their local workspace. 



\subsection{Perspective-sharing: Environment Modifications}
\label{modificationEnvironment}
\begin{figure}[t]
    \centering
    \includegraphics[width=1\linewidth]{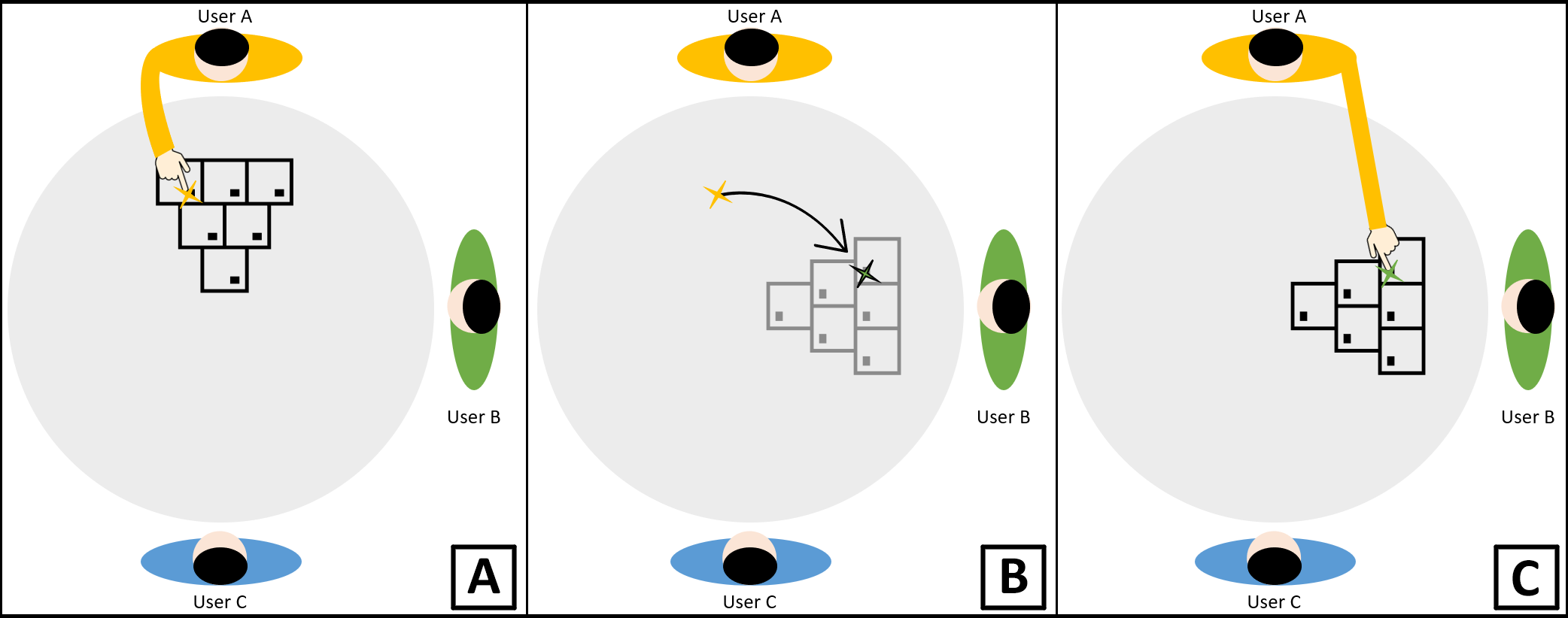}
    \caption{Transformation applied to the avatar's arm and hand position: [A] - User A points at a point in his workspace, [B] - User B receives the information of where user A pointed and maps it to his own workspace, [C] - User A's opposing arm is rendered as a spline and their hand points at the mapped point in User B's workspace.}
    \label{fig:transformations_pointing}
\end{figure}

To obtain a common perspective over the workspace, manipulations are done to the local environment of each user. These are based on the seat that the user takes around the table. The workspace is divided into eight increments of 45 degrees, starting from the Assembler position, i.e., the right of the table in a top view. In Fig.~\ref{fig:transformations_pointing}, there is an example. In [A], we see user A's POV, which has the workspace rotated by 90 degrees counterclockwise. In [C], representing user B's POV, their workspace does not require any rotation since they are at position 0, i.e., right of the table. User C will need a rotation of 270 degrees since they are at position 7 in the table, while B is in position 1 (0 degrees) and A is in position 3 (90 degrees). In a setting with eight users, the space between User A and User B would have another user with an increment of 45 degrees and a user between User C and A with an increment of 315 degrees. 

\subsection{Perspective-sharing: Avatar Modifications}
\label{modificationAvatar}
\begin{figure*}[t]
    \centering
    \includegraphics[width=.96\textwidth]{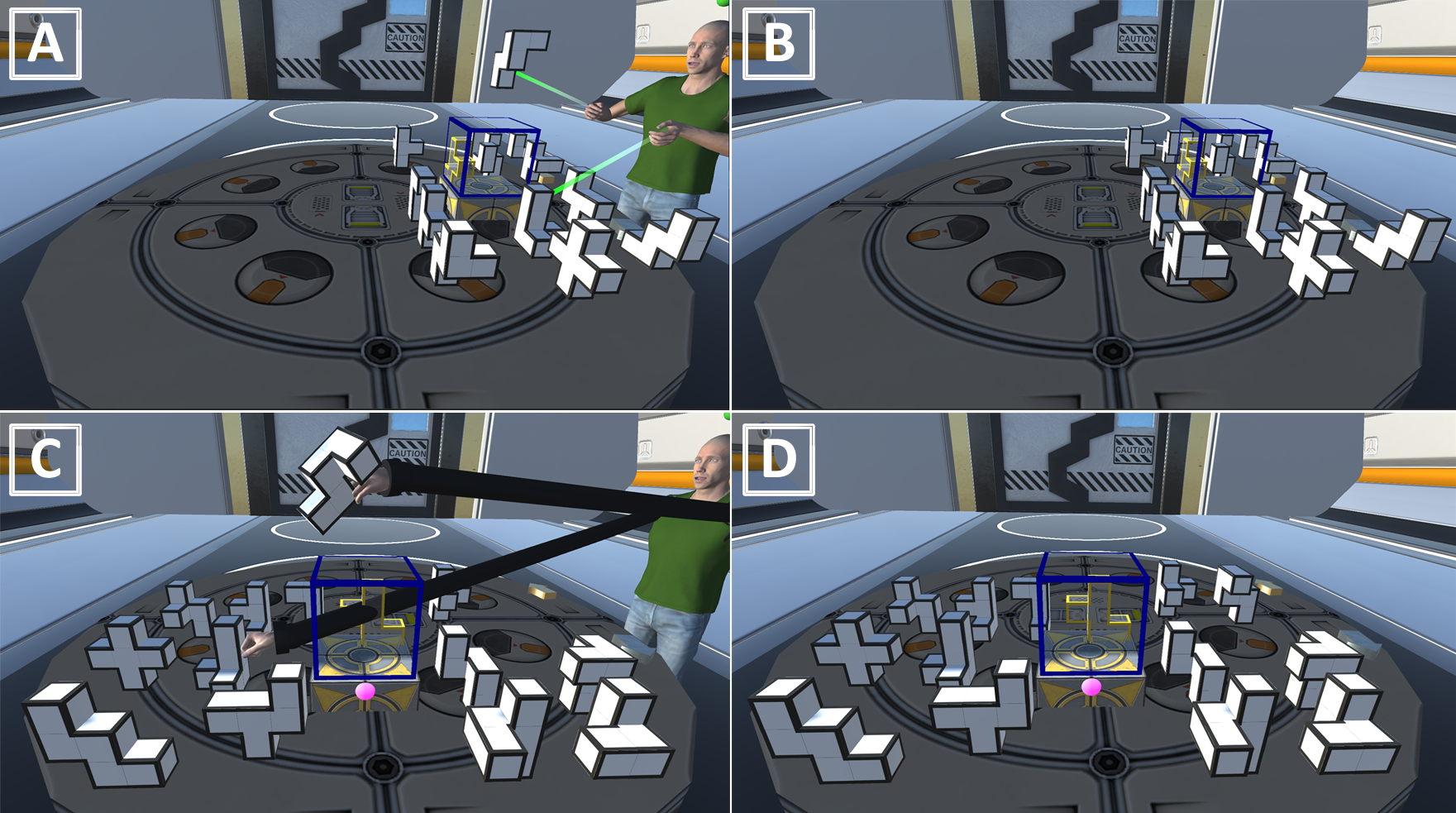}
    \caption{Different conditions in the same POV from the perspective of Instructor 1: [A] and [B] represent the Veridical condition. [C] and [D] represent SPARC}
    \label{fig:condition_veridical_vs_suri}
\end{figure*}

In our POC, there are two modes of interacting and referencing the workspace: by direct interaction, using their virtual hands, or indirect interaction, using their virtual rays or pointers (see Fig.\ref{fig:condition_veridical_vs_suri}).

Due to the locally applied rotation to the user's workspace, the meaning of the deictic gestures and any referencing of the workspace is lost. To counteract this, our approach maps the references made by each user in their workspace to the local user's workspace. This is done in three main steps that are shown in Fig.~\ref{fig:transformations_pointing}: [A] - User A points at a point in their local workspace with their right hand; [B] - User B receives the point where user A's right hand is and maps it to their workspace (by rotating it by the difference in rotation between the two users, i.e., 90\textdegree) [C] - User B renders user A's left hand to the mapped point with a line connecting to it.

In our implementation, the manipulation of the avatars does not always happen. It will only occur if the user is referencing something within the workspace. The trigger will happen if the hand enters the workspace or the ray points at an object inside the workspace. This trigger occurs for each hand individually. 

Each user is assigned an angle, and this is used to calculate the others' deictic references. The information regarding each user's hand position is propagated and then rotated around the center point, i.e., the center of the table, by the difference in degrees between the local user (LU) and the remote user (RU). To  connect the RU's avatar to these new calculated references, we used Splines 
that are rendered in real-time in the LU's environment coming from the RU's arm to their hand positioned at the calculated reference, as shown in Fig.\ref{fig:avatar_manipulation_final}, where the angle difference will be 180\textdegree. Notice that there was a mirroring of the RU. This is done to avoid the arms crossing over one another when both hands are being used to reference something.




%
\section{Evaluation}
\label{chap:evaluation}

We designed and conducted a user study to assess the effects of the SPARC on task performance, workspace awareness, and sense of co-presence in the VE. Our POC demonstrated to be technically capable of supporting up to eight users at a time. Nevertheless, for the sake of simplicity, we chose to test with trios only. Thus, any effect involving a different number of co-workers is out of the scope of this assessment. With trios, we address a scenario that is challenging to the previous works with a shared perspective. We tested two conditions:

\begin{itemize}
    \item \textbf{Condition V}: Veridical collaboration. Participants stood around a virtual table and collaborated with different individual perspectives of the task, as it would be in real-life interactions.
    \item \textbf{Condition S}: Collaboration under SPARC. Participants stood around a virtual table with a shared perspective and avatar manipulations active.
\end{itemize}

\subsection{Task}
We chose a 3D assembly task represented by the Bedlam Cube. It is a 3D puzzle with thirteen distinct pieces designed to fit together in a 4x4x4 frame. With over 19,000 combinations, the assembler can't guess which piece fits each position. In our setting, each participant in a trio is assigned one of three roles: Assembler, Instructor 1, or Instructor 2. The assembler is the one who can manipulate objects. The two instructors have access to the instructions, i.e., the solution for the next piece, but cannot manipulate the pieces. The instructors see the outline of a piece correctly placed inside the cube and must indicate to the assembler which piece and where it should go. In such a scenario, the two instructors have to communicate with each other and with the assembler to share their knowledge, which must be understood by the assembler, who manipulates the pieces, one at a time, until the task is completed with the assembly of the full cube.

An example of the puzzle in the POV of an instructor is depicted in Fig.~\ref{fig:condition_veridical_vs_suri}D, where the yellow outline inside the empty cube represents the piece and its orientation. A grid-snapping system is employed to aid in the positioning of the pieces. When a piece is dropped or rotated, its pose will be corrected to this grid - this removes issues regarding overly precise positioning. 

Users are constrained to life-like perspectives in the veridical condition (V). The puzzle is directed towards the assembler, and the instructors have a side view of the puzzle. The instructor on the left-hand side of the assembler will see the puzzle from the left of the puzzle. A comparison between the two conditions can be seen in Fig.~\ref{fig:condition_veridical_vs_suri}, where the POV of Instructor 1 is seen under the two conditions.

\subsection{Procedure}
Participants began by completing a user profile and a consent form. Next, they were provided with an explanation of how to use the HMD and interact in the POC. The task was also explained at this point, and participants decided who would be assigned to each role, with the task and roles further elaborated on as needed. After adjusting their HMD, participants familiarized themselves with their respective roles and the different types of interactions available in the VE. In particular, the assembler was instructed on moving and rotating the pieces using direct and indirect interaction modes. 

The design is within subjects. The 18 participants are grouped into 6 trios, and each trio performs the task once in condition V and once in condition S. The order between conditions is counterbalanced.
The participants began in one of the conditions and were tasked with completing one puzzle until all the pieces were correctly placed. Following this, there was a short break during which participants completed a trial questionnaire. The pieces were then jumbled, and the task was repeated under the other conditions. Another trial questionnaire was completed afterwards.

\subsection{Measurements}
We collected the following data:

\begin{enumerate}
    
\item \textbf{Total completion time}: End time minus initial time, in seconds. 
\item \textbf{Errors per piece}: Counted when a piece is released at a wrong position within the cube (it returns to its place on the table).
\item \textbf{Number of attempts}: Counted when a piece is grabbed and released outside of the cube.
\item \textbf{Eye-contact time}: Time spent looking at the other players. Computed when a raycast from the center point of the avatar's head intercepts another avatar's head.

\end{enumerate}

\noindent These other metrics are derived from the measures above:

\begin{enumerate}
    
\item \textbf{Total errors}: The sum of the errors per piece.
\item \textbf{Total moves}: The sum of errors and attempts. Note that the successful moves are not accounted for as they are forcefully the same for every completed puzzle, i.e., 13.

\end{enumerate}

To evaluate user preference and satisfaction, we applied a questionnaire that compiled questions that evaluate various aspects of a collaborative task in a VE: the perceived workload of the task, the sense of co-presence felt, and user awareness. It was composed mostly of statements evaluated on a 6-point Likert scale. At the end of the questionnaire, the users could also give their thoughts on the task and report any bugs they experienced.

\subsection{Participants}
The subject group comprised 18 participants, with 8 females and 10 males. Their ages ranged from 20 to 39, with an average age of approximately 24 years (M = 24, SD = 4.14). All participants had attained a university education.

Eight individuals reported using videoconference platforms, such as Skype, Zoom, or Microsoft Teams, at least once a day. Six participants used these platforms at least once a week, while four reported rare usage. Concerning VR environments, the data indicated that most participants (15 out of 18) had never used VR environments, and the remaining three participants reported infrequent use.

\section{Results}

\begin{figure*}[t]
    \centering
    \includegraphics[width=.96\textwidth]{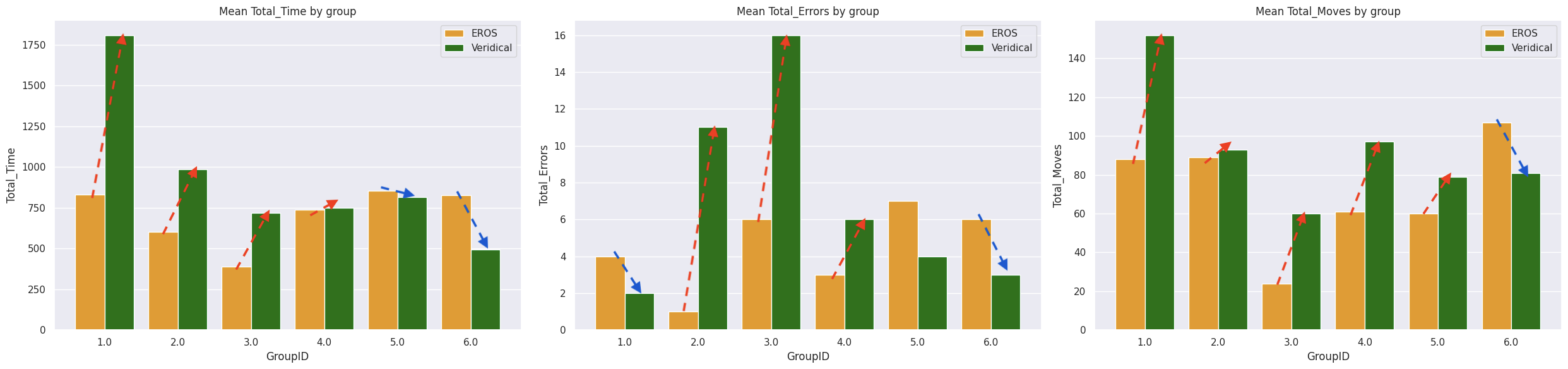}
    \caption{Raw data from the experiment depicting the three measures variation by trios for the two conditions. The lower, the better. Arrows indicate the direction of the difference for each group: red when veridical is higher and blue when SPARC is higher.}
    \label{fig:raw data}
\end{figure*}

Fig.~\ref{fig:raw data} shows the raw data collected from the experiment. In a preliminary analysis, we looked for tendencies and eventual outliers. Groups 1-3 performed condition V first, and groups 4-5 started with condition S. Two facts are strikingly noticeable: (1) green bars (condition V) tend to be higher than orange ones (condition S) for groups 1 to 3, and the trend inverts for the remaining groups; (2) group 6 is the only one with an inverted trend in the three measures. These facts suggest an order effect. On the one hand, V-first presents higher times, errors, and moves for V than S.
On the other hand, S-first has more mixed results, with fairly lower S bars than groups 1-3, which did not improve much in condition V. Our interpretation is that V is harder for everyone but becomes easier after some experience with S. The opposite is not true. The S-performing groups first perform well and continue to perform well in V, with only slight improvement. Concerning Group 6, it presented a pattern that was diverse enough to be excluded from the analysis due to its profile.

A third factor is also noticeable. Group 1 presented much higher times and moves than the average of the other groups. However, we identified that this group spent all that additional time and moves playing with a single piece. However, we did not remove that piece from the analysis because we did not record individual times per piece.

For further analysis, we kept the data from 15 participants, i.e., five groups and ten trials. Overall statistics are in Table~\ref{tab:overall_results}. 

\begin{table}[t]
\centering
\caption{Descriptive statistics}
\label{tab:overall_results}
\begin{tabular}{lrrr}
          & \multicolumn{3}{c}{\textbf{(Mean $\pm$~\textbf{SD})}} \\ \cline{2-4} 
\multicolumn{1}{c}{\textbf{}} & \textbf{Total Time} & \textbf{Total Errors} & \textbf{Total Moves} \\ \hline
SPARC      & 681 $\pm$~192    & 4.2 $\pm$~2.3  & 64 $\pm$~26  \\
Veridical & 1015 $\pm$~455   & 7.8 $\pm$~5.6  & 96 $\pm$~34  \\
Total     & 848 $\pm$~373    & 6 $\pm$~4.5    & 80 $\pm$~33  \\ \hline
\end{tabular}
\end{table}

We detail the results in the following, grouped into four different categories. It is worth noticing that with five groups, the statistical power of our sample is not strong. Yet, as the results will show, there are statistically significant findings, and non-significant data also have interesting trends.

\subsection{User preferences}

\begin{figure*}[t]
    \centering
    \includegraphics[width=.9\textwidth]{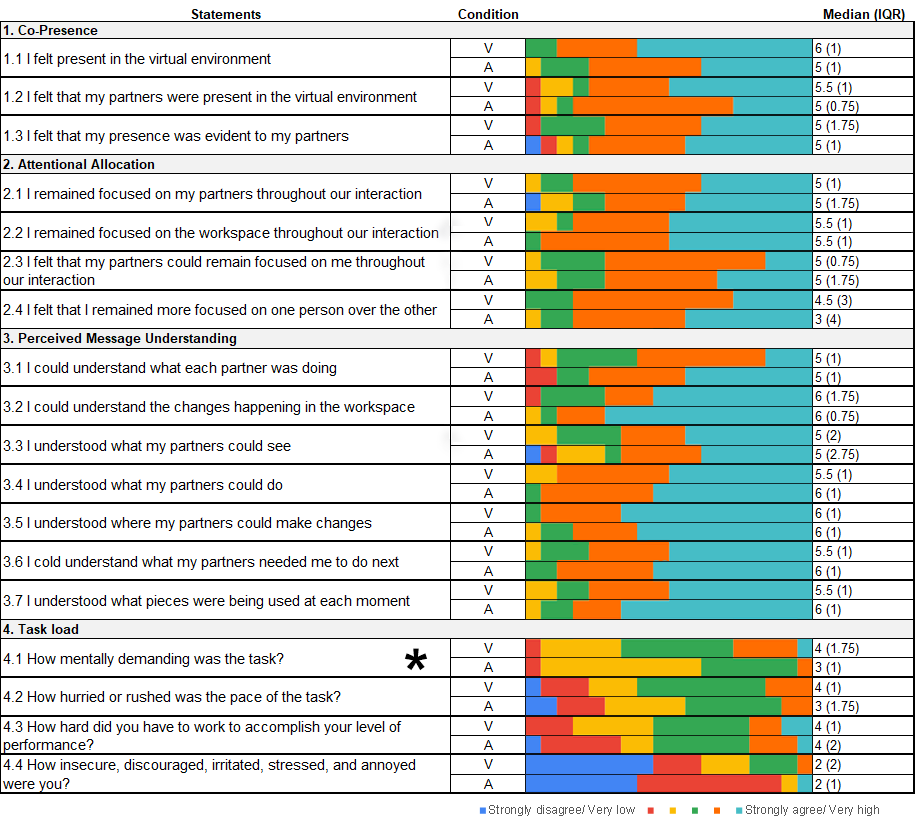}
    \caption{Results from the user preferences questionnaire for both our approach(A) and Veridical(V) conditions. * indicates statistical significance.}
    \label{fig:user_preference_graph}
\end{figure*}

The applied questionnaires and associated results are summarized in Fig.~\ref{fig:user_preference_graph}. We employed the Wilcoxon Signed Ranks test to check for the significance of the differences. Most of them showed no statistical significance. This hints that co-presence, attentional allocation, and perceived message understanding are weakly affected by the tested conditions. 

In terms of Task Load, the test reported a significant increase in mental demand in the Veridical condition ("How mentally demanding was the task?"), where \textit{Z} = -2,486, \textit{p} = .013.

\subsection{Task performance}

After running the Shapiro-Wilk test on time, errors, and total moves, we could assume a normal distribution for the data. We then used a paired T-test to check for the significance of the differences.

\begin{figure}[t]
  \begin{subfigure}[t]{.32\textwidth}
    \centering
    \includegraphics[width=\linewidth]{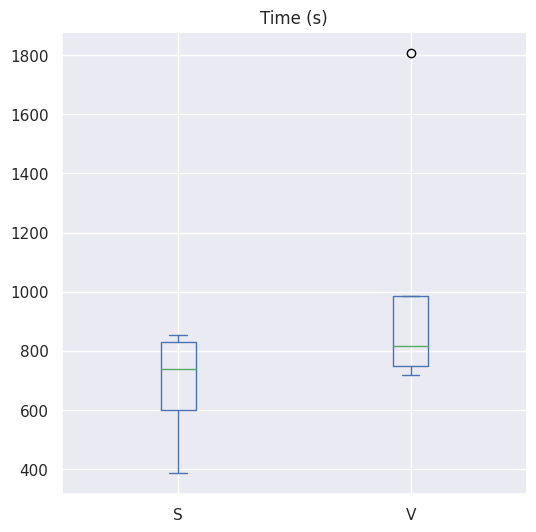}
    \caption{Total time}
    \label{fig:total_time}
  \end{subfigure}
  \begin{subfigure}[t]{.315\textwidth}
    \centering
    \includegraphics[width=\linewidth]{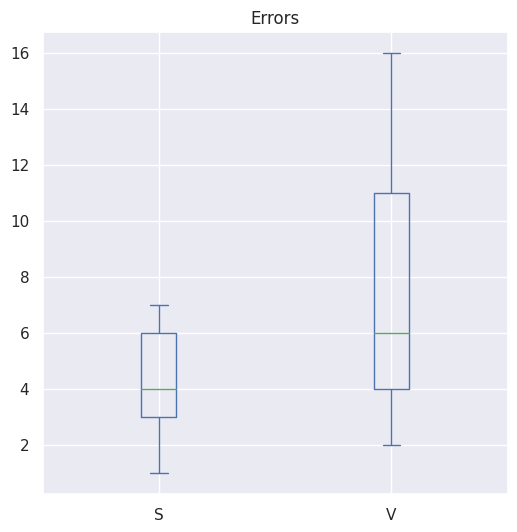}
    \caption{Total errors}
    \label{fig:total_errors}
  \end{subfigure}
  \begin{subfigure}[t]{.32\textwidth}
    \centering
    \includegraphics[width=\linewidth]{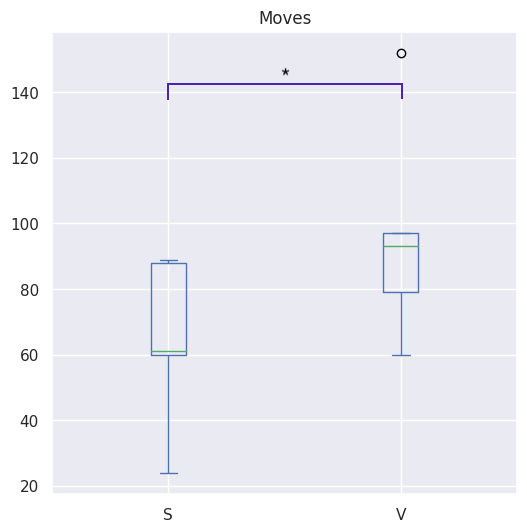}
    \caption{Total moves (movement economy)}
    \label{fig:move_economy_plot}
  \end{subfigure}

    \caption{Summary of the experimental data analysis comparing SPARC (S) and Veridical (V) conditions.}
    \label{fig:quantitative results}
\end{figure}

There was no statistically significant difference in the total times with results t(6) = -1.83, \textit{p}= .14 between SPARC (\textit{M} = 681, \textit{SD} = 192) and the Veridical approach (\textit{M} = 1015, \textit{SD} = 455). We attribute the lack of significance to the small sample size. However, the average number of times suggests an advantage for SPARC regarding time efficiency. Fig.~\ref{fig:total_time} illustrates the magnitude of this potential advantage.

The mean results for the total number of errors are SPARC (\textit{M} = 4.20, \textit{SD} = 2.38) and veridical (\textit{M} = 7.80, \textit{SD} = 5.67). Again, we performed a Paired T-Test (t(6) = -1.28, \textit{p} = .26), which also resulted in no statistical significance for the difference of the means. See also Fig.~\ref{fig:total_errors}. 

We analyzed the movement economy as the number of times pieces were picked up until the puzzle's completion.The results for the total moves between our approach (\textit{M} = 64.4, \textit{SD} = 26.5) and veridical (\textit{M} = 96.2, \textit{SD} = 34.3) are depicted in Fig.~\ref{fig:move_economy_plot}. We performed the Paired T-Test and, this time, obtained statistical significance with t(6) = -3.17, \textit{p} = .033.

We observed that in the Veridical condition, users with the role of instructor tended to move around the table and reposition themselves beside the assembler, possibly to have a better view of the pieces or the puzzle. About the observed communication at the moments of discovering which piece was going on the cube, users mostly verbalized references to their pointers or hands (e.g.\textit{"this piece"}). We also observed that explaining rotations was harder under SPARC for some users. In general, some interactions were also limited because the other user's arms were in the way of the user's workspace. 

\section{Conclusions and future work}

We introduced SPARC to tackle workspace occlusions in collaborative interactions from a shared perspective. It features a common perspective while preserving the meaning of deictic references. Users can then maintain a common understanding of the task space and how others interact with it. Using a round-table setting, all users can see each other, which means they can understand and be understood by others relative to their intentions and nonverbal cues. They do this without shifting their attention between task and person spaces.

The reports of our user study concluded that collaborating in a shared perspective meant lower mental demand than in the veridical condition. It could be argued that the awareness of the workspace was improved in SPARC by creating a common task, reference, and personal spaces, without requiring users to shift focus between them, which would be supported by lower mental demand. Moreover, a mix of significant and non-significant quantitative results indicate that SPARC helps improve an assembly task's performance.

In the future, we aim to conduct a larger user study with more simultaneous users in the VE, where we will explore the effects of different numbers of users. Scaling to more users would also create new challenges, such as the overload of arms stretching in the workspace and the difficulty of understanding where each user is by where their voice originates. 


\section*{Acknowledgements}

This work was supported by national funds through FCT, \textit{Fundação para a Ciência e a Tecnologia}, under grant 2022.09212.PTDC (XAVIER Project), and project UIDB/50021/2020 (DOI:10.54499/UIDB/50021/2020) under the auspices of the UNESCO Chair on AI\&VR of the University of~Lisbon.

\bibliographystyle{splncs04}
\bibliography{Mendeley} 

\end{document}